# Ground state of the competing spin chain $Cs_2Cu_2Mo_3O_{12}$

Yukihiro HOSHINO, Soichiro ATARASHI, Takayuki GOTO*, Masashi HASE[1] and Takahiko SASAKI[2]

*Physics Division, Sophia Univerisity, Chiyodaku, Tokyo 102-8554, Japan*
*[1] National Institute for Materials Science (NIMS), Tsukuba, Ibaraki 305-0047, Japan*
*[2] Institute for Materials Research, Tohoku University, Sendai 980-8577, Japan*

*E-mail: gotoo-t@sophia.ac.jp*

The high-field ground state of the competing-spin-chain compound $Cs_2Cu_2Mo_3O_{12}$ with the ferromagnetic first-nearest-neighbor $J_1$ = –93 K and the antiferromagnetic second-nearest-neighbor $J_2$ = +33 K was investigated by $^{133}$Cs-NMR. A divergence of $T_1^{-1}$ and a peak-splitting in spectra were observed at $T_N$ = 1.85 K, indicating the existence of a field-induced long range magnetic order. In the paramagnetic region above 4 K, $T_1^{-1}$ showed a power-law temperature dependence $T^{2K-1}$. The exponent $K$ was strongly field-dependent, suggesting the possibility of the spin-nematic Tomonaga Luttinger Liquid state.

**KEYWORDS:** quantum spin system, NMR, antiferromagnet, spiral structure

## 1. Introduction

Frustrated spin chains are simple models that still give us surprisingly rich physics. The quasi-one dimensional spin chain compound $Cs_2Cu_2Mo_3O_{12}$ possesses two dominant exchange interactions along the chain, the ferromagnetic first-nearest-neighbor $J_1$ = –93 K and the antiferromagnetic second-nearest-neighbor $J_2$ = +33 K, determined from the detailed measurements of the uniform magnetic susceptibility and magnetization under high fields [1-3]. The ground state of this compound attracts much interest recently, because competing interactions make this compound to be a good candidate for the spin nematic phase, which is considered to be a two-magnon condensated state [4-6]. The spin nematic derives its name from the inversion symmetry in the $S_z$ = 0 state of $S$ = 1 spin, which is tends to be formed in the present system due to the ferromagnetic exchange interaction $J_1$ between the nearest-neighboring two spins.

Theoretical study on the ground state of the competing spin chain system of this type has been performed by Hikihara *et al.* [7]. They have built up the phase diagram for wide range of the coupling ratio $J_1/J_2$ and predicted the appearance of the nematic or multipolar Tomonaga-Luttinger-liquid (TLL) spin state under high fields[7-10]. Sato *et al.* has discussed the effect of inter-chain interactions to competition between the nematic oder and the antiferromagnetic order in finite temperatures [11].

Experimentally, it has so far been reported by one of authors M. H. that the present system has a gapless spin state under zero field, the saturation field of approximately 10 T, and no evidence of the phase transition down to 2 K [1,2]. In this paper, we investigate microscopically by NMR technique the ground state of this system under high applied fields between 4 and 8 T. The observed $^{133}$Cs-NMR spectra and the longitudinal nuclear spin relaxation rate $T_1^{-1}$ show an evidence of the magnetic order at low temperature below 2 K and also the possibility of TLL spin liquid state above its critical temperature.

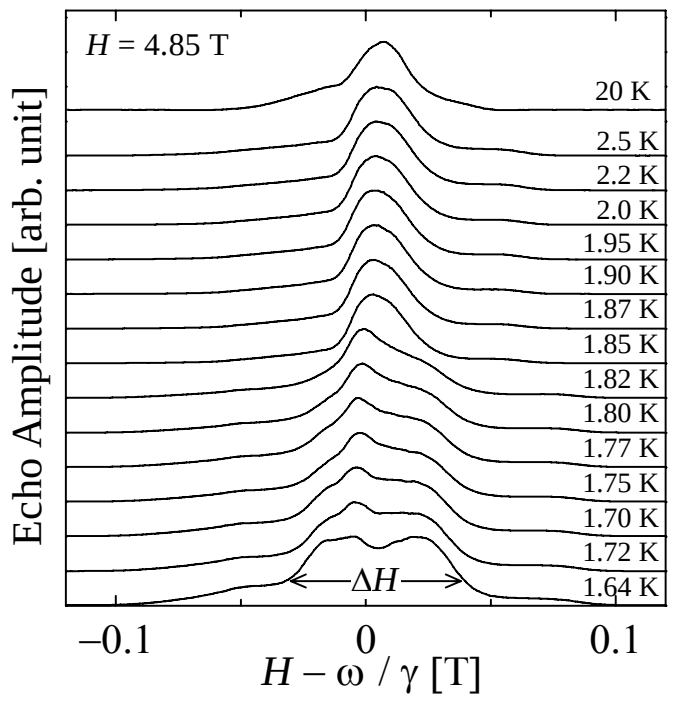

**Fig. 1.** The profile of NMR spectra at various temperatures. The field-sweep region is between 4.71 and 5 T. The definition of the line width $\Delta H$ is shown bv an arrow.

## 2. Experimental

The powder sample of $CsCu_2Mo_3O_{12}$ was synthesized by the solid-state reaction method [1,2]. The system has a monoclinic crystal symmetry, which is isomorphic to $RbCu_2Mo_3O_{12}$ [3] and belongs to the space group of *C*2/*c*, with a zig-zag chain of Cu, so called as a ribbon chain, running along *b*-axis. There are three inequivalent Cs sites, 4*d*, 4*e* and 8*f* positions, locating approximately at the interstitial position of the chains. The distances to the nearest neighboring Cu sites are between 4 and 6 Å, which is expected to give a moderate intensity of hyperfine interaction.

NMR measurements were performed in the temperature range between 1.6 and 20 K. The spectra were obtained by recording the amplitude of spin-echo against the applied field [12,13]. The longitudinal spin relaxation rate $T_1^{-1}$ was obtained by the conventional saturation-recovery method with a pulse-train for saturation. The time evolution of the nuclear spin magnetization after saturation was traced until the difference to the thermal equilibrium value becomes one or two percent, and fitted to the stretched exponential function of $1 - \exp[-(\tau/T_1)^{\beta}]$, where $\tau$ is the duration after the saturation, to obtain the temperature dependent two constants $T_1$ and $\beta$. The latter is an index related to the inhomogeneity in the system; in general, $\beta$ decreases from unity, the pure case value, as the system becomes inhomogeneous [14].

In general, the relaxation curve for $^{133}$Cs nuclear spin with $I$ = 7/2 should show the so-called multi exponential function.[15] However, in the present case, the electric quadruple moment of $^{133}$Cs is very small, that is, less than two percent of $^{63}$Cu, so that all the signals that come from the satellite and central transitions of the three inequivalent sites merge and become a single reservoir, which however contains many different microscopic states in it, and hence is considered to be inhomogeneous. This explains why the observed nuclear spin relaxation curves were stretched exponential type, and also is consistent with the fact that the spectrum in the paramagnetic state was a single peak as will be stated below.

## 3. Results

Figure 1 shows the typical profile of spectra of powder sample. At high temperatures, we only observed a single symmetric peak centered nearly at the zero shift position and broad tails at both sides. This comes simply from the fact that the electric quadruple interaction in $^{133}$Cs nuclei is negligibly small as state above so that the satellite and central transitions of the three inequivalent sites merge to make a single peak. As lowering temperature, the peak changed its shape abruptly to an asymmetric one below $T_N$ = 1.85 K, and then gradually broadened until it split into two peaks at the lowest temperature.

The temperature dependence of the NMR line width $\Delta H$ is shown in Fig. 2. It starts increasing at $T_N$ with the temperature dependence of $|T_N - T|^{0.5}$, the critical exponent of which agrees with the molecular field theory.

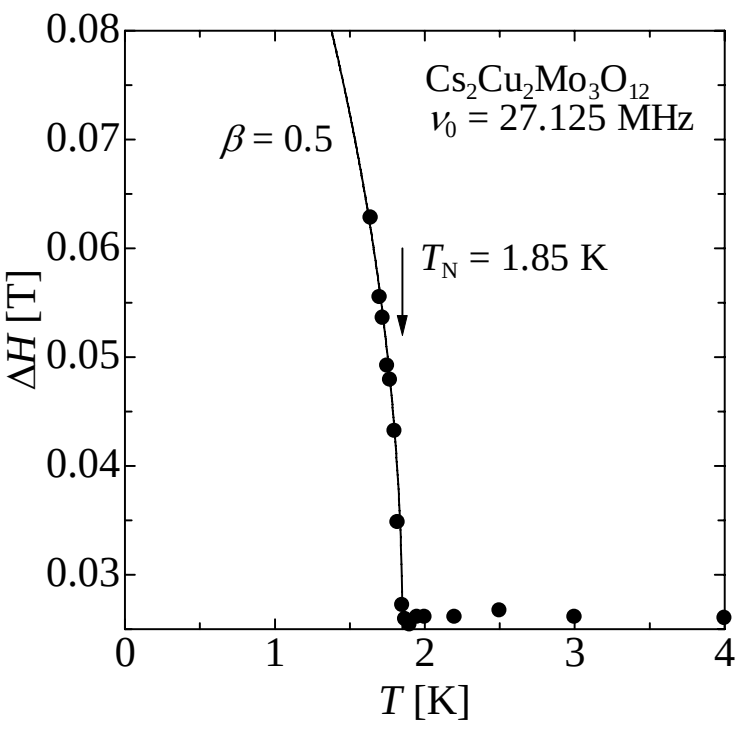

**Fig. 2.** The temperature dependence of the line width $\Delta H$. The solid curve shows the power-law function $|T_N – T|^{0.5}$.

Figure 3 shows the temperature dependence of $T_1^{-1}$. In high temperature region, it gradually increased with decreasing temperature and showed a critical divergence at around $T_N$, which agrees with the one determined by spectra. The index of the stretched exponential function $\beta$ stayed nearly constant at around 0.4 and gradually decreased to 0.25 below $T_N$. [14]

The Néel temperature $T_N$ was field dependent; it took a maximum of 1.85 K at around 4.85 T, while 1.7 K at 4.35 T and below 1.6 K at 5.75 T. The field 4.85 T corresponds to the point where the uniform magnetization is approximately a half of its saturation value. Though this suggests that the observed magnetic order is field-induced, more detailed measurements in a wide field region are required for a definitive argument, and are now in progress [16,17].

Next, we proceed to the field dependence of $T_1$ in the paramagnetic region. Its temperature dependence measured at various fields within 4 – 8 T is shown in Fig. 4. In this temperature region between 4 and 20K, which is far apart from $T_N$, $T_1^{-1}$ obeys a power law of temperature $T^{2K-1}$, where the index $K$ increases from 0.05 to 0.93 as increasing the applied field. Note that the gradient of the $T_1^{-1}$ in the log-log plot obviously changes its sign when $K$ passes through 0.5, which corresponds to the field approximately 7 T. Thus obtained $K$ is plotted against the uniform magnetization $M$ in Fig. 5 [2]. It monotonically increases with increasing $M$ toward the saturation value.

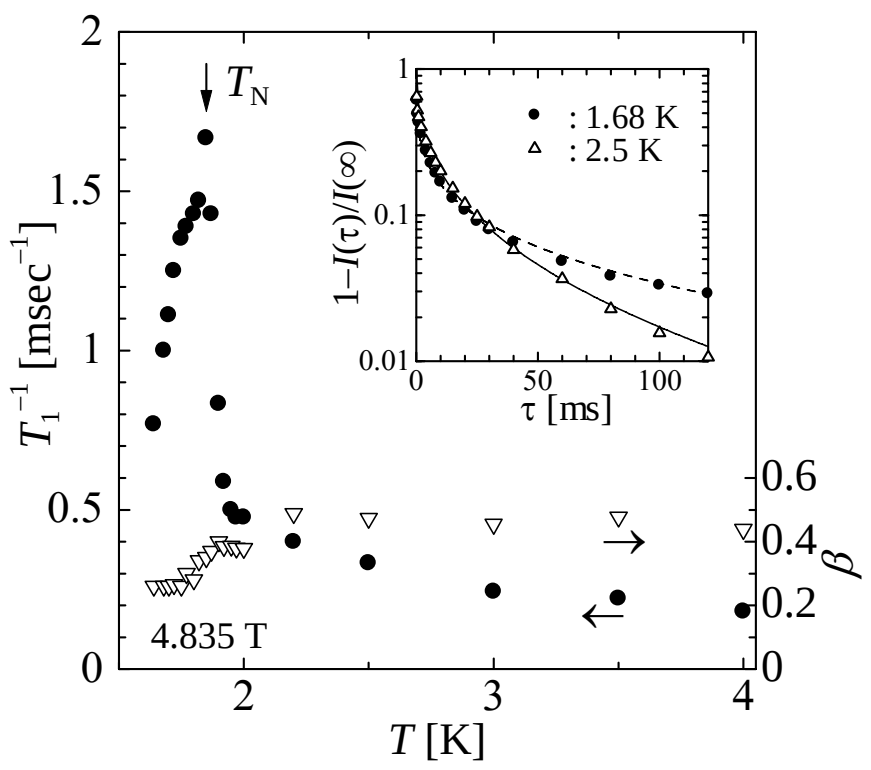

**Fig. 3.** The temperature dependence of the longitudinal nuclear spin relaxation rate $T_1^{-1}$ and the index $\beta$ in the stretched exponential function in the vicinity of $T_N$. The inset shows typical profile of nuclear spin relaxation curves.

## 3. Discussion

The broadening and splitting in the spectrum at low temperature indicate an emergence of static staggered component among the 3$d$-spin moments, and hence demonstrate the existence of magnetic order. The idea of magnetic order is further supported by the temperature dependence of $T_1^{-1}$ in Fig. 3, which shows an existence of critical slowing down due to a second order thermodynamic phase transition. The apparent observation of the static hyperfine field justifies that this is a conventional magnetic order rather than a nematic order, although the latter may be more appealing. For the present system is quasi one dimensional, observed three dimensional long-range order is considered to be brought by the inter-chain spin interaction[11,18].

The characteristic split pattern of the spectrum at lowest temperatures suggests the magnetic structure with a specific propagation vector $q$. However, at present , it is difficult to determine $q$ from the powder signal, and the measurement on a single crystal is strongly urged.

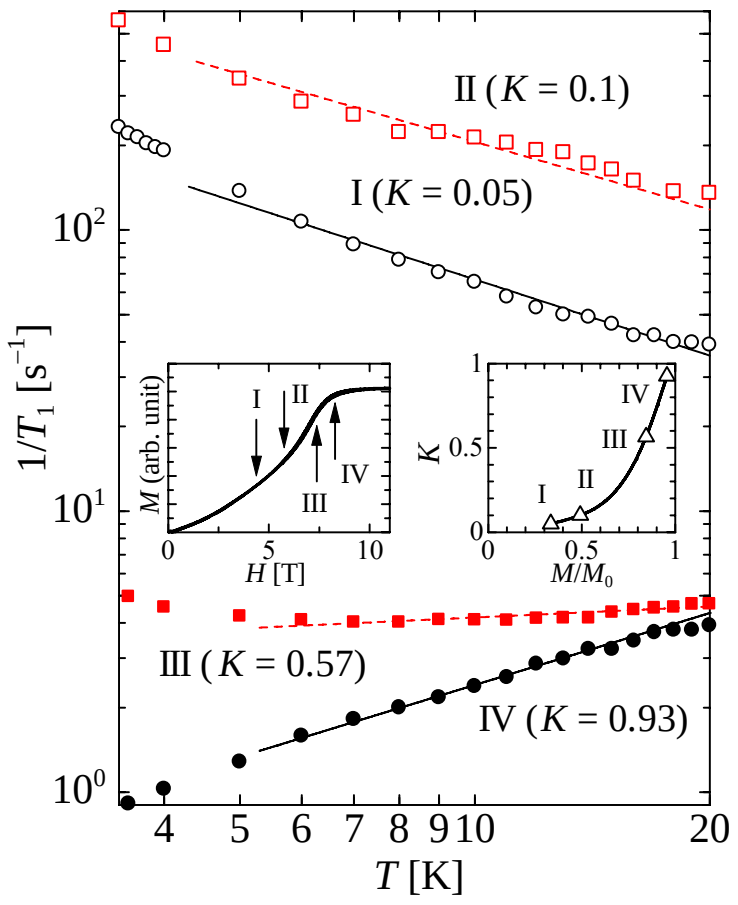

**Fig. 4.** The temperature dependence of $T_1^{-1}$ under fields 4.38, 5.75T, 7.37 and 8.28 T shown as I - IV, respectively. The fitted lines show the power-law function of temperature, $T^{2K-1}$, where $K$ is the Luttinger paramter. The insets show (left) the $M$-$H$ curve with the measured field positions, and (right) the Luttinger parameter $K$ (=½$\eta$) versus uniform magnetization $M$ scaled by its saturation value $M_0$.

Next, we compare the temperature dependence of $T_1^{-1}$ in different magnetic fields with the theory on the nematic TLL state [19,20]. In general, $T_1^{-1}$ is approximately proportional to the longitudinal and transverse electron spin correlation function via the diagonal and off-diagonal components in the hyperfine coupling tensor, respectively. In the conventional TLL phase, both the longitudinal and transiverse spin correlation functions decay with the power law as $x^{-2K}$, where $x$ is the distance, and $K$ (= ½$\eta$) is the Luttinger parameter [8-10,19,20]. The contributions to $T_1^{-1}$ from these two sectors contribute are $T^{2K-1}$ and $T^{1/2K-1}$, respectively, and hence at low temperatures, $T_1^{-1}$ should always increase with decreasing temperatures, irrespective of the value of $K$ except for ½ [19,20].

Recently, it has been shown by Sato *et al.* that in the nematic or multipolar TLL phase, decay of the transverse correlation function is cut off exponentially, because of the necessity of breaking a magnon bound state in order to create an excitation with $\Delta S_z = \pm 1$. Consequently, the term $T^{1/2K-1}$ vanishes and the temperature dependence of $T_1^{-1}$ is expected to change drastically at $K$ = ½, that is, from the positive gradient for $K$ < ½ to the negative gradient for $K$ > ½. The observed $T_1^{-1}$ shown in Fig. 4 just matches the theoretical prediction, and is consistent the idea of the emergence of the nematic or multipolar TLL state in this system. This is also supported by the behavior of $K$ against uniform magnetization $M$ shown in the inset of Fig. 4. The monotonic increase in $K$ with $M$ is compatible to the recent theoretical calculation by Hikihara [7]. For quantitative and determinant argument, more data on a still wider field range and a determination of the hyperfine coupling tensor are inevitable, and are now in progress.

## 4. Conclusion

We have measured NMR spectra and $T_1^{-1}$ on the competing spin chain system

$Cs_2Cu_2Mo_3O_{12}$ under wide range of magnetic fields. At intermediate field region 4.85 T, we have observed the long range magnetic order at $T_N$ =1.85 K, which was field dependent, suggesting that the order is field-induced. In paramagnetic temperature region, $T_1^{-1}$ obeyed the power law temperature dependence $T^{2K-1}$. The index $K$ showed significant increase with increasing magnetic field and hence $M$, the behavior of which is consistent with the theoretical prediction on the nematic TLL state.

## Acknowledgment

This work was partially supported by JSPS KAKENHI Grant Number 21110518, 24540350 and 21540344. The NMR measurements were performed in part at High Field Laboratory for Superconducting Materials, Institute for Materials Research, Tohoku University.